\documentclass[
    prapplied, twocolumn, superscriptaddress, nofootinbib,
    amsmath, amssymb, aps, floatfix, preprintnumbers,10pt
]{revtex4-2}

\usepackage[utf8]{inputenc}
\usepackage{setspace}
\usepackage[dvips]{graphicx}
\usepackage[dvipsnames]{xcolor}
\definecolor{CiteBlue}{RGB}{45,52,151}
\usepackage{booktabs,tabularx}
\usepackage{siunitx}
\usepackage{makecell}
\usepackage[
    colorlinks=true,
    linkcolor=CiteBlue,
    urlcolor=CiteBlue,
    citecolor=CiteBlue
]{hyperref}

\begin{document}
\title{Thermal Analog Computing: Application to Matrix-vector Multiplication with Inverse-designed Metastructures}
\author{Caio Silva}
\affiliation{Department of Physics, Massachusetts Institute of Technology, Cambridge, Massachusetts 02139, USA}

\author{Giuseppe Romano}
\email{romanog@mit.edu}
\affiliation{Institute for Soldier Nanotechnologies, Massachusetts Institute of Technology, 77 Massachusetts Avenue, Cambridge 02139, MA, USA}
\date{\today}
\begin{abstract}

The rising computational demand of modern workloads has renewed interest in energy-efficient paradigms such as neuromorphic and analog computing. A fundamental operation in these systems is matrix–vector multiplication (MVM), ubiquitous in signal processing and machine learning. Here, we demonstrate MVM using inverse-designed metastructures that exploit heat conduction as the signal carrier. The proposed approach is based on a generalization of effective thermal conductivity to systems with multiple input and output ports: The input signal is encoded as a set of applied temperatures, while the output is represented by the power collected at designated terminals. The metastructures are obtained via density-based topology optimization, enabled by a differentiable thermal transport solver and automatic differentiation, achieving an accuracy $>99\%$ in most cases across a pool of matrices with dimensions $2\times2$ and $3\times3$. We apply this methodology---termed thermal analog computing---to realize matrices relevant to practical tasks, including the discrete Fourier transform and convolutional filters. 
These findings open new avenues for analog information processing in thermally active environments, including temperature-gradient sensing in microelectronics and thermal control systems.

\end{abstract}

\maketitle
\pagestyle{myheadings}
\markboth{Authors}{Thermal Analog Computing: Application to Matrix-vector Multiplication with Inverse-designed Metastructures}

\section{Introduction}

Modern classical computing is built upon binary logic, where transistors act as the main building blocks that toggle between discrete “0” and “1” states. This paradigm has powered decades of progress, but it also imposes fundamental limits: every operation requires digital switching and repeated memory access. As a result, these processes dissipate energy as heat due to resistive losses~\cite{pop2006heat}. As computational workloads continue to scale, there is growing motivation to explore computing paradigms that harness the intrinsic physical behavior of materials rather than relying on Boolean abstractions~\cite{analogreviwew2025}. A key primitive in nearly every analog platform is matrix–vector multiplication (MVM), the backbone of signal processing, control systems, and machine learning~\cite{lecun2015deep}. Many analog computing platforms have been engineered to perform MVMs, including switch networks~\cite{Sun2022AMC}, memristor arrays~\cite{song2024programming}, and photonic metamaterials~\cite{silva2014performing, cordaro2023solving, Nikkhah2024}. A comprehensive overview of these efforts is provided in the review by Zangeneh-Nejad et al.~\cite{zangeneh2021analogue}. One natural carrier of information remains underexplored: heat. The field of thermal metamaterials—born from the framework of transformation thermodynamics~\cite{guenneau2012transformation}—has yielded striking demonstrations such as cloaks, rotators, and concentrators~\cite{li2021transforming}. However, most of these advances focused on guiding or shaping heat flow, rather than computing with it.  We propose a new paradigm for MVM using inverse-designed thermal metastructures, where heat is not a byproduct of computation—it is the \emph{signal} itself. 
In this framework, which we refer to as \emph{thermal analog computing}, input vectors are encoded as temperature fields applied to metal ports, and output powers are collected as the resulting heat fluxes. The challenge, then, is geometric:
\emph{what structure should conduct heat such that its response encodes a desired matrix transformation?} We tackle this problem using density-based topology optimization~\cite{sigmund2013topology}, a technique that discretizes a material into pixels of variable density. Each pixel has the ability to continuously transition between “solid” and “void,” enabling the optimizer to adjust the geometry until the desired input-output relationship is realized. In our framework, each matrix is represented by a collection of structures that are optimized simultaneously. The thicknesses of these structures also serve as degrees of freedom, and they are optimized in conjunction with the material densities. Our computational framework is based on an in-house differentiable thermal solver (pending release) built using JAX~\cite{jax2018github}, enabling a fully differentiable end-to-end pipeline. We employ this framework to matrices with increasing complexity, showcasing practical applications such as the Fast-Fourier Transform and the Convolution filter (Toeplitz matrix). In all examples, which include matrices of dimensions $2\times2$ and $3\times3$, we consistently achieve exceptional fidelity, with accuracy $> 99\%$ in most cases. 

Our method is fundamentally different from proposals based on thermal logic gates, which have long been studied in analogy to electronics~\cite{li2012colloquium,wong2021review}. Classic examples include thermal diodes~\cite{chang2006solid,liang2009acoustic,hamed2019thermal} and transistors~\cite{li2006negative,lo2008thermal}. However, these devices operate on discrete logic states—“hot” and “cold”—and thus mirror digital computing. In contrast, our approach enables calculations in the \emph{continuous} regime, directly using heat flow as the information carrier. Consequently, it circumvents the need for repeated digital-to-analog conversion and logic switches. Our approach potentially opens the door to in-situ computation in environments where heat naturally exists: on-chip sensing of temperature gradients~\cite{gradiometer}, thermal regulators~\cite{park2024development}, and distributed temperature sensing and mapping~\cite{bucher2022printed}.
\\

The paper is structured as follows. In Sec.~\ref{sec:methodology}, we lay down the methodology, including the thermal model and the optimization pipeline. Then, in Sec.~\ref{sec:single} and Sec.~\ref{sec:multiple}, we showcase our method for single- and multiple-structure designs, respectively. In Sec.~\ref{sec:applications}, we showcase our method for matrices of practical application. In Sec.~\ref{sec:perfomances}, we summarize the performances of our structures, and in Sec.~\ref{sec:ballistic} we analyze the validity of Fourier's law in the proposed systems. In Sec.~\ref{discussions}, we highlight the limitations and future outlook of our work, followed by Conclusions (Sec.~\ref{sec:conclusions}). We include several Appendices that further elaborate on various elements of our methodology, such as the cost function (Sec.~\ref{sec:optimization}), bandwidth (Sec.~\ref{sec:bandwidth}), energy efficiency (Sec.~\ref{sec:energy}), signal-to-noise ratio (Sec.~\ref{sec:snr}), and nonlinear thermal effects (Sec.~\ref{sec:nonlinear}). 

\section{Methodology}\label{sec:methodology}

\begin{figure*}[t]
    \includegraphics[width=\linewidth]{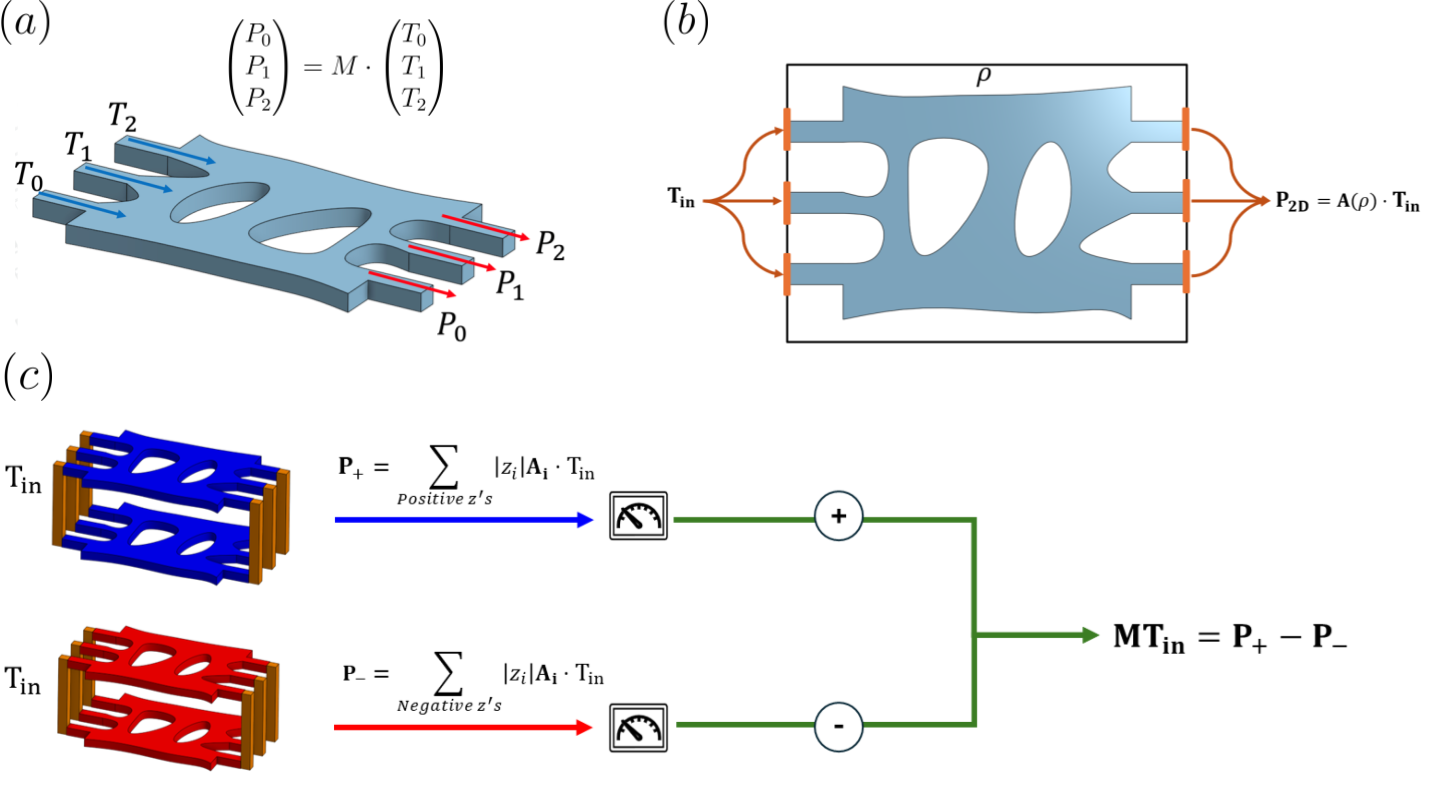}
    \caption{
    (a) Input temperatures $\mathbf{T}_{\mathrm{in}}$ are applied through thermalized ports, and output powers $\mathbf{P}_{\mathrm{out}}$ are collected from the opposite side.
    For linear, temperature-independent conductivities, the steady-state relation 
    $\mathbf{P}_{\mathrm{out}} = \mathbf{M}\mathbf{T}_{\mathrm{in}}$ 
    defines a conductance matrix $\mathbf{M}$ that depends solely on the geometry and material distribution. 
    (b) In simulations, each metastructure is represented as a binary 2D map $\rho(x,y)$, where $\rho=1$ denotes conductive material and $\rho=0$ denotes vacuum. 
    The simulated 2D power response $\mathbf{P}_{\mathrm{2D}} = \mathbf{A}(\rho)\mathbf{T}_{\mathrm{in}}$ defines a geometry-dependent operator $\mathbf{A}(\rho)$, which is later scaled by the wafer width $z$ to obtain the 3D conductance matrix. 
    (c) Because thermal conduction cannot produce negative flux, we represent arbitrary real-valued matrices as the weighted sum of physically realizable positive components. 
    The proposed infrastructure in a realistic setting is that structures positive $z_i$ (blue) and negative $z_i$ (red) are operated separately; their respective outputs, $\mathbf{P}_+$ and $\mathbf{P}_-$, are then digitally subtracted to yield the effective analog operation 
    $\mathbf{M}\mathbf{T}_{\mathrm{in}} = \mathbf{P}_+ - \mathbf{P}_-$.
    }
    \label{fig:MVM}
\end{figure*}

Our goal is to perform matrix–vector multiplications (MVMs) through heat conduction. To this end, we generalize the effective conductance to the effective conductance \emph{matrix}, where input vectors are encoded as temperature fields applied to the metastructure through thermalized metal ports 
\(\mathbf{T}_{\mathrm{in}} \in \mathbb{R}^{N_{\mathrm{in}}}\).  
The output is obtained as the thermal power collected at the output ports,  
\(\mathbf{P}_{\mathrm{out}} \in \mathbb{R}^{N_{\mathrm{out}}}\).  
All output ports are kept at the reference temperature $T_0$, ensuring that no net current flows when all input temperatures are $\mathbf{T}_{\mathrm{in}} = T_0$. (As we assume linear regime, in our simulation we can safely set $T_0$ = 0 K.)  Assuming a temperature-independent conductivity, Fourier’s law implies a linear relationship between input temperatures and output powers:
\begin{equation}\label{linearequation}
    \mathbf{P}_{\mathrm{out}} = \mathbf{M}\, \mathbf{T}_{\mathrm{in}},
\end{equation}
where the matrix \(\mathbf{M} \in \mathbb{R}^{N_{\mathrm{out}} \times N_{\mathrm{in}}}\) has units of W/K and depends solely on the geometry and material distribution of the structure.  
The matrix $\mathbf{M}$ can be regarded as the conductance operator mapping the input vectors into the output vectors. This relationship, visualized in Fig. \ref{fig:MVM}-a, allows analog computation of $\mathbf{y} = \mathbf{M}\mathbf{x}$, where $\mathbf{x}$ is the input temperature and $\mathbf{y}$ is the output power. Similarly to the effective thermal conductivity~\cite{Romano2012}, the matrix $\mathbf{M}$ can be tuned by the geometry of our device. Thus, our task is to solve the inverse problem:  
Given a target matrix \(\mathbf{M}_{\mathrm{target}}\), find the structure whose conductance matrix best approximates it. In our framework, we envision the physical device as a two-dimensional (2D) metastructure with conductance matrix $\mathbf{A}$ and thickness $L_{z}$, giving
\begin{equation}\label{2d:MVM}
    \mathbf{M} = L_z \mathbf{A}.
\end{equation}
The structure yielding the matrix conductance $\mathbf{A}$ is obtained by density-based topology optimization~\cite{sigmund2013topology} (Sec.~\ref{sec:topopt}), where a material is described by the density $\rho(x,y)$ (Fig.\ref{fig:MVM}(b)).
One critical limit, however, is that because heat conduction obeys the laws of thermodynamics, negative matrix elements cannot be realized. This limitation is not unique to our approach. Analog computing systems based on memristive circuits also face similar restrictions: electrical conductances can only be positive. A common workaround is to represent a target matrix as the difference between two nonnegative matrices~\cite{uysal2021xmap,ielmini2018inmemory},
\begin{equation}\nonumber
    \mathbf{M} = \mathbf{M}_+ - \mathbf {M}_-,
\end{equation}
where both \(\mathbf{M}_+\) and \(\mathbf{M}_-\) contain only positive entries.  
Each submatrix is realized by a separate physical device, and the final result is recovered digitally by subtracting the two measured outputs. Inspired by this idea, we generalize our framework to decompose any desired matrix into \(K\) physically realizable components,
\begin{equation}\label{eq:MM}
    \mathbf{M} = \sum_{i=0}^{K-1} z_i\, \mathbf{A}(\rho_i).
\end{equation}
In Eq.~\ref{eq:MM}, each \(\rho_i\) corresponds to distint 2D structures with thickness \(|z_i|\equiv L_{z_i}\). Crucially, the coefficients \(z_i\) are not restricted in sign, so Eq.~\ref{eq:MM} can accommodate both positive and negative contributions. In practice, we perform computations on structures for positive and negative \(z_i\) separatly and perform digital post-subtraction of their resulting cumulative powers (Fig.\ref{fig:MVM}(c)).

After discretizing the densities into $N$ pixels, our inverse design problem is then described in terms of the parameters $\{\boldsymbol \rho_i\}_{i=1}^{K}$ and coefficients \(\boldsymbol{z} =\{z_i\}_{i=1}^{K}\). 
The total number of degrees of freedom (DOFs) is
\begin{equation}\nonumber
    N_{\text{DOFs}} = (N + 1)K,
\end{equation}
where \(N\) is the number of pixels that define each structure. The following sections detail the underlying heat conduction model and the optimization pipeline used to solve this PDE-constrained inverse-design problem.

\subsection{Thermal Model}
Thermal transport is calculated by the heat conduction equation, which, for a given input vector $\mathbf{T}$, reads
\begin{eqnarray}\label{eq:fourier}
\nabla \cdot \kappa \nabla T = 0, & \quad & \mathbf{x} \in \Omega, \\
T = T_i^{\text{in}}, & \quad & \mathbf{x} \in \Gamma^{\text{in}}_i, \text{ for } i = 0, \ldots, M-1 \nonumber \\
T = 0, & \quad & \mathbf{x} \in \Gamma^{\text{out}}_j, \text{ for } j = 0, \ldots, N-1 \nonumber \\
\kappa \nabla T \cdot \mathbf{n} = 0, & \quad & \mathbf{x} \in \partial \Omega \setminus (\Gamma^{\text{in}} \cup \Gamma^{\text{out}}) \nonumber ,
\end{eqnarray}
where $\kappa(\mathbf{x})$ is the space-dependent thermal conductivity; the term $\Omega$ is the computational domain with size $L$ = 12 cm, $\partial \Omega$ being its outer boundary.  The input and output contacts are denoted by $\Gamma^{\text{in}}_i$ and $\Gamma^{\text{out}}_j$, respectively. The remainder of the boundary is treated as a thermal insulator. Equation~\ref{eq:fourier} is implemented in a JAX-based~\cite{jax2018github} differentiable finite-volume technique, which will be documented elsewhere. The solver has been recently validated against ANSYS for non-homogeneous systems~\cite{romano2025diffchip}. The grid consists of $N \times N$  (pixels), with $N = 100$. Once Eq.~\ref{eq:fourier} is solved, the output power is computed by
\begin{equation}\label{eq:flux2d}
P_{\text{out},j} = -\kappa\int_{\Gamma^{\text{out}}_j} \nabla T \cdot \mathbf{\hat{n}}\mathbf{x}.
\end{equation}
We note that Eq.~\ref{eq:flux2d} give the power per unit of length. The actual power output is obtained after multiplying $\textbf{P}_{\text{out}}$ by $|z_i|$ as described in the previous section.

\subsection{Topology Optimization}\label{sec:topopt}
 Inverse design is performed by density-based topology optimization within the three-field approach~\cite{sigmund2013topology}; the optimization is performed in terms of a fictitious density $\rho \in [0,1]$ (here, for simplicity, we consider its continuous representation), which is allowed to continuously vary between the void (0) and the solid phase (1). To avoid checkerboard patterns, $\rho$, known as the \emph{design density}, is filtered using the conic kernel
\begin{align} \nonumber 
 w(\mathbf{x}) &= 
 \begin{cases} 
 \frac{3}{\pi \tilde R^2}\left(1-\frac{|\mathbf{x}|}{\tilde R}\right), & \quad |\mathbf{x}| < \tilde R \\ 
 0, & \quad \text{otherwise},
 \end{cases} 
\end{align}
with conic radius $R$ = 8 $\mu$m. The \emph{filtered field}, $\tilde \rho = w\ast\rho $, is then projected to obtain a binary optimal structure so that it can be manufactured. A common choice for this operation is the 
tanh projection
\begin{equation}\label{eq:tanh}
\hat \rho =  \frac{\tanh{\beta \eta} + \tanh\left(\beta\left(\tilde{\rho}-\eta\right)\right)}{\tanh{\beta \eta}+ \tanh\left(\beta\left(1-\eta\right)\right)},
\end{equation}
where the $\beta$ parameter tunes the strength of the projection. A common strategy is to increase $\beta$ as optimization progresses so that it can initially explore various topologies. One issue with this approach, however, is that for high beta the problem becomes near-nondifferentiable, slowing down, or even stalling the optimization. For this reason, we instead use the recently introduced subpixel-smoothed projection (SSP)~\cite{hammond_unifying_2025}, which solves this issue and allows for smooth differentiation up to $\beta=\infty$. Choosing the $\beta$ schedule is problem dependent as it is a trade off between computational efficiency and optimized structure quality. In this work, we adopt the $\beta$-schedule: $\beta = [2,4,16,32,64,128,256,\infty]$, where the each epoch is run for 60 iterations and the remaining one for 30. The projection field is then used in the heat conduction equation, precisely by mapping it into a thermal conductivity tensor via
\begin{equation}\label{eq:local}
\kappa(\mathbf{x}) = \left[\epsilon +\kappa_0\left(\hat \rho(\mathbf{x})-\epsilon\right)\right]\mathcal{I},
\end{equation}
where $\kappa_0$ is the thermal conductivity of Si (150 Wm$^{-1}$K$^{-1}$ at $T_0 = 300$ K). We regularize Eq.~\ref{eq:local} with the parameter $\epsilon=10^{-4}$ to avoid numerical instability due to regions where $\hat \rho = 0$.

\subsection{Optimization Framework}
\begin{figure}[t]
    \includegraphics[width=1\linewidth]{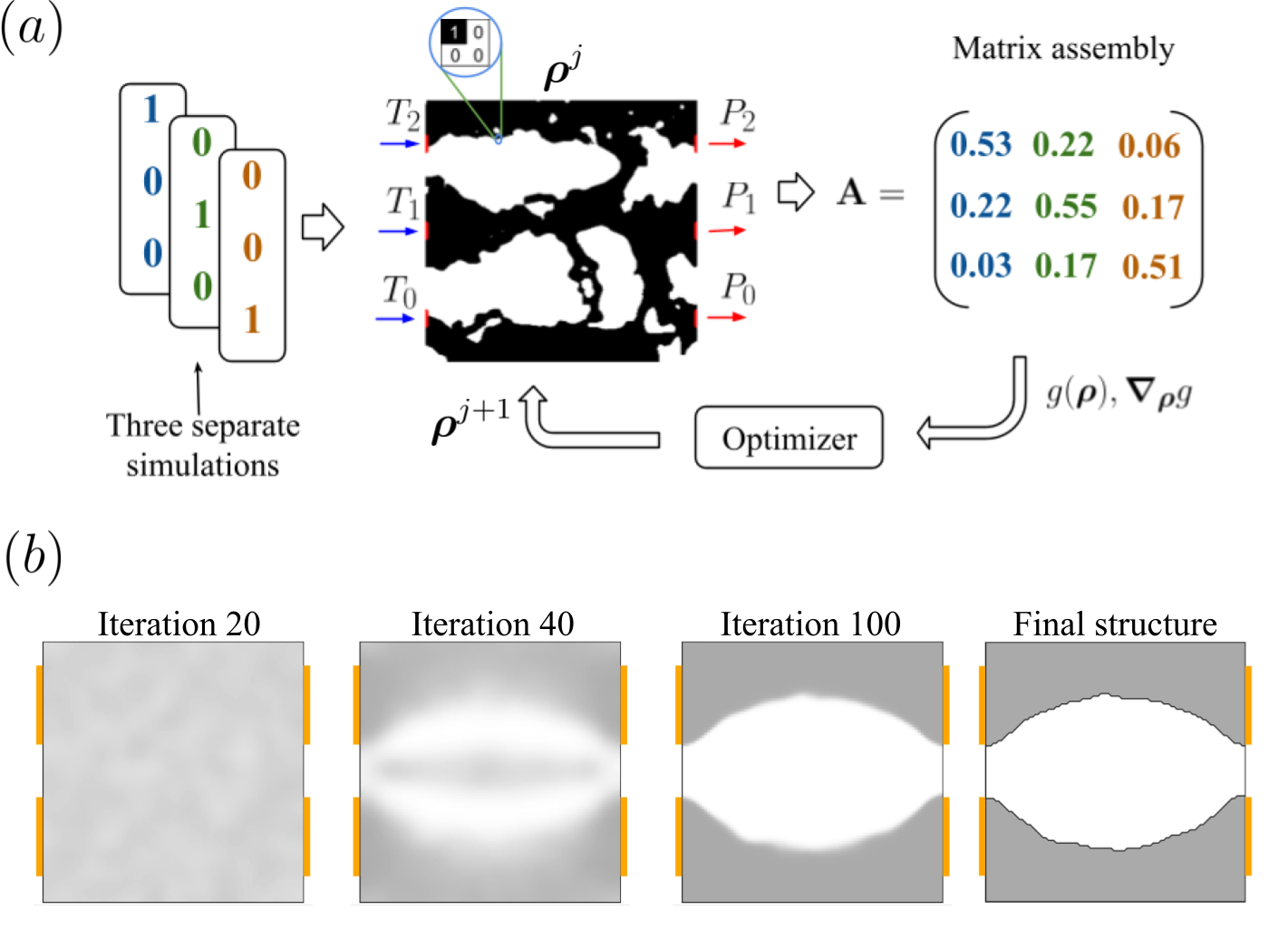}
    \caption{(a) Overview of the optimization workflow for a 3$\times$3 thermal metastructure performing matrix--vector multiplication (MVM).
The device geometry is parameterized by the material density field $\rho$, where $\rho = 1$ and $\rho = 0$ correspond to solid and void regions, respectively.
At each iteration, three steady-state heat conduction simulations are performed using orthogonal temperature inputs $\mathbf{T}_0$, $\mathbf{T}_1$, and $\mathbf{T}_2$ (blue arrows), representing the basis vectors of the input space.
The corresponding output powers $\mathbf{P}_0$, $\mathbf{P}_1$, and $\mathbf{P}_2$ (red arrows) form the columns of the 2D conductance matrix $\mathbf{A}$, which is used to calculate $\mathbf{M}$. From the reconstructed matrix, the cost function is calculated.
The optimizer updates the density field $\boldsymbol{\rho}$ using the computed objective function $g(\rho)$ and its gradient $\nabla_{\rho} g$, generating the next iteration $\boldsymbol{\rho}_{i+1}$. 
(b) Evolution of the density field during optimization, showing the emergence of the final topology. Intermediate structures gradually converge toward a well-defined heat-guiding geometry, where conductive paths connect the thermal ports (orange).}
    \label{OptimizerPipeline}
\end{figure}

Our optimization framework is formulated in terms of the densities $\boldsymbol \rho_i$ and effective thicknesses $z_i$, with $k=1...K$, which we collectively describe as $\boldsymbol\theta$. To reconstruct the matrix $\mathbf{M}$ for a given set of parameters, we perform a set of $D$ simulations ($D$ being the dimensionality of the matrix) where we activate only one input port by applying $1K$. As this is the expansion of the canonical basis, the submatrices $\mathbf{A}_i$ are simply obtained by stacking the resulting power outputs as column vectors (Fig.~\ref{OptimizerPipeline}). The matrix $\mathbf{M}$ is then computed by Eq.~\ref{eq:MM}. The cost function has two building blocks; the first is 
\begin{equation}\label{eq:L2}
    \mathcal{L}_2 = \frac{\|\mathbf{W} \odot (\mathbf{M}(\boldsymbol\theta) - \mathbf{M}_\text{target})\|_2}{\sqrt{N_\text{in} N_\text{out}}},
\end{equation}
which is the $L_2$ norm, weighted by the matrix $\mathbf{W}$. This matrix balances the contribution of entries with different magnitudes and is defined as
\begin{equation} 
W_{ij} =
\begin{cases}
\dfrac{1}{|M^\text{target}_{ij}|}, & |M^\text{target}_{ij}| > 0, \\[10pt]
\dfrac{N_\text{in} N_\text{out}}{\|\mathbf{M}_\text{target}\|_F}, & |M^\text{target}_{ij}| = 0,
\end{cases}
\label{eq:Wdef}
\end{equation}
where $\|\cdot\|_F$ denotes the Frobenius norm of the matrix. This matrix  ensures that each term in the loss contributes approximately equally on a relative scale: larger entries are not overrepresented, and smaller entries retain non-negligible optimization pressure. The second line of Eq.~\ref{eq:Wdef} is defined to avoid a singularity for zero entries. In early iterations, the optimizer often “breaks” heat pathways—leading to matrix elements that undershoot their targets (becoming too small). Rebuilding such paths later is difficult, while reducing an overshoot (too large) is relatively easy. To bias optimization toward maintaining connectivity, we add a contribution to the cost function $\mathcal{B}$, which is detailed in Appendix A. The total cost function is then $g = \mathcal L_2 + \mathcal B$. Even when $g$ is minimized, the final structure might barely touch the input ports, having most of the current going through a few pixels, introducing instability and unrealistic conduction patterns. To prevent this, we introduce a constraint on the maximum value of the magnitude of the heat flux, normalized by its average value, $(\max |\mathbf{J}|)/\bar{|\mathbf{J}|}$. Since the maximum value is not a differentiable function, we regularize it via the p-norm
\begin{equation}\nonumber
    \mathcal{P} = \left(\sum_\Omega \tilde{J}_b(\mathbf{r})^p \, d\mathbf{r} \right)^{1/p},
\end{equation}
with $p=20$. Throughout this work, we choose the threshold value $R^{\text{thresh}} = 15$. Our optimization framework reads
\begin{align}\label{eq:algo}
\min_{\boldsymbol\theta}\quad & g(\boldsymbol\theta) \\
\text{s.t.}\quad & \frac{\mathcal{P}}{\overline{|\mathbf{J}|}} - R^{\text{thresh}} \le 0 \notag \\
\text{s.t.}\quad & 0 \le \rho_{in} \le 1 \quad \text{n=1...N}, \quad \text{i=1...K} \notag\\
\text{s.t.}\quad & |z_i| \le z_{\text{max}} \quad \text{i=1...K}, \notag
\end{align}
We define \(z_\text{max}\) as the characteristic scale of the device thickness required to sustain fluxes on the order of \(\|\mathbf{M}_{\mathrm{target}}\|_F\). Balancing the target magnitude with the nominal conductance of a port of width \(w\) and length \(L\) in a medium of conductivity \(\kappa_0\) gives
\begin{equation}\label{z0}
    z_{\text{max}} \;=\; 3\,\frac{\|\mathbf{M}_{\mathrm{target}}\|_F}{\kappa_0\, w/L}\, .
\end{equation}
In practice, \(z_{\text{max}}\) is used to scale (and bound) the thickness variable \(z\), which discourages the optimizer from collapsing the geometry into few-pixel filaments and then compensating with unrealistically large \(z\). As the optimizer, we employ the Conservative Convex Separable Approximation (CCSA)~\cite{svanberg1987method}, implemented in Nlopt~\cite{johnson2014nlopt}. A schematic of the optimization is shown in Fig.~\ref{OptimizerPipeline}.
\\
The optimization algorithm described in Eq.~\ref{eq:algo} is executed sequentially for each value of $\beta$, using the optimized solution from the previous step as the initial guess for the next. The $\beta$-schedule, defined in Sec.~\ref{sec:topopt}, terminates with $\beta=\infty$. Within the SSP formalism~\cite{hammond_unifying_2025}, fully projected fields are \emph{quasi-binary}, meaning that they become strictly binary in the continuum limit $\Delta x \rightarrow 0$. In the discrete setting, this corresponds to the presence of a \emph{subpixel} boundary layer of nonbinary material along each interface. Because manufacturable materials must be strictly binary, we postprocess the final design through full binarization, which introduces a small change in the response function. To compensate for this effect, we rescale the parameter $z_k$ such that the norm of the matrix $z_k\mathbf{A}_k$ matches its value \emph{prior} to binarization. In practice, this correction is on the order of $1\%$. All reported values of $\mathbf{z}$ and error metrics in this work refer to fully binarized structures.

\begin{table}[t]
\centering
\footnotesize
\setlength{\arraycolsep}{4pt}
\renewcommand{\arraystretch}{1.1}
\begin{tabular}{@{} l c @{}}
\toprule
\textbf{Type of Matrix} & \textbf{Error ($\%$)} \\
\midrule
2×2 Identity & 0.71 \\
2×2 Directional & 10$^{-5}$ \\
2×2 Two-Branch & 0.68 \\
2×2 Straight-Path & 3.82 \\
2×2 Positive Off-Diagonal (K=2) & 0.78 \\
2×2 Signed Off-Diagonal (K=3) & 10$^{-5}$ \\
2×2 Large Magnitude Contrast (K=4) & 1.2 \\
3×3 General Coupled (K=4) & 2.93 \\
2×2 Hadamard & 0.5 \\
2×2 Rotational ($\theta=\pi/3$, K=3) & 3.09 \\
3×3 Rotational Toeplitz (K=3) & 1.87 \\
3×3 DFT (Real part) & 3.55 \\
3×3 DFT (Imag part) & 1.13 \\
\bottomrule 
\end{tabular}\caption{Error for each matrix computed with Eq.~\ref{eq:L2}.}\label{table:Accuracy}
\end{table}

\section{Single-structure Designs}\label{sec:single}

\begin{figure*}[t]
    \centering
    \includegraphics[width=0.80\linewidth]{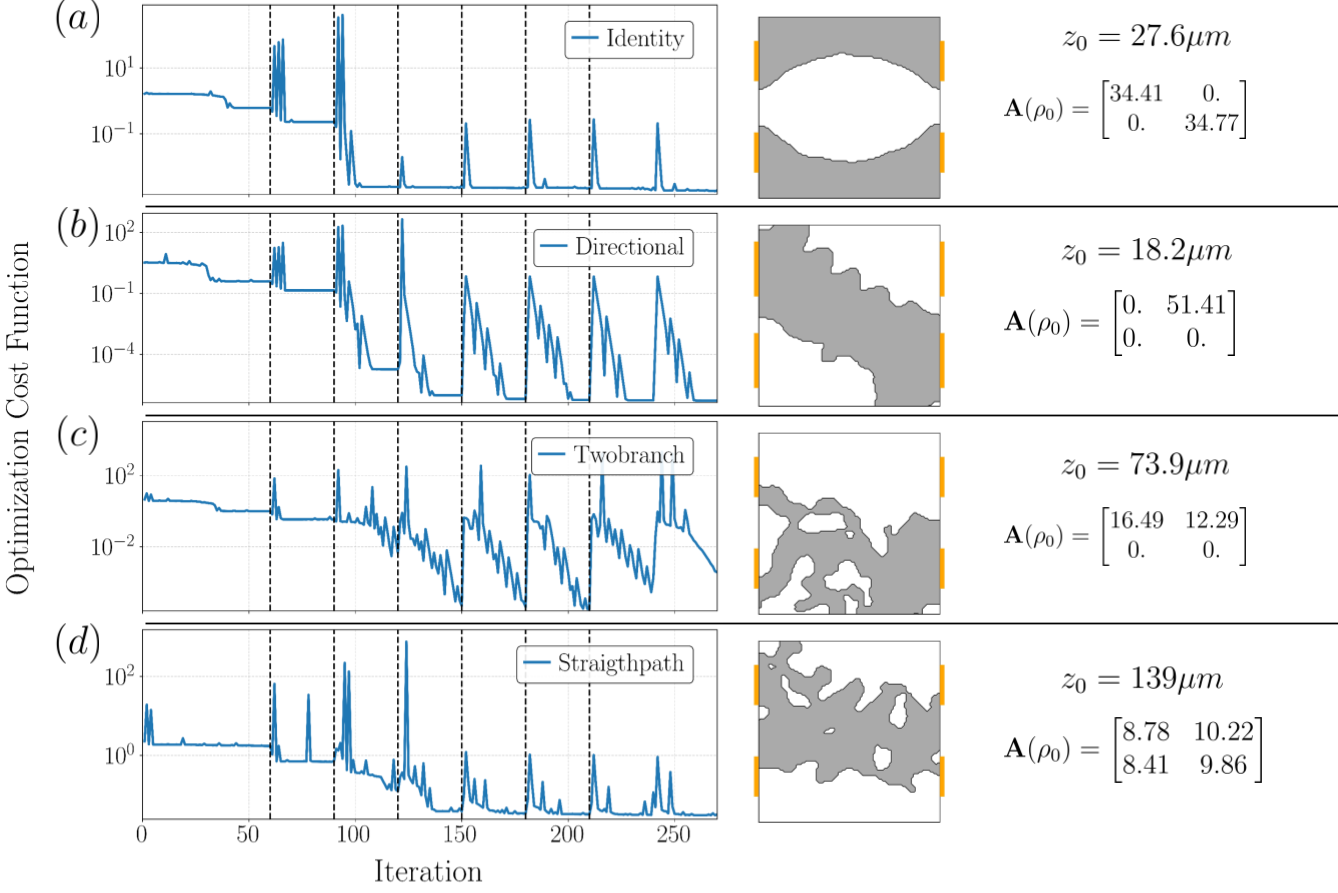}
    \caption{
    Optimization trajectories and final inverse-designed structures for four target matrices: a)Identity, b) Directional, c) Two-branch, and d) StraightPath.  Each subplot shows the evolution of the cost function over optimization iterations.  Vertical dashed lines indicate updates of the regularization parameter $\beta$. The right panels display the final material distributions $\rho(\mathbf{r})$ after convergence. For each design, the effective propagation length $z_0$ and final response matrix $\mathbf{A}(\rho_0)$ are reported, representing the physical length and two-dimensional transfer matrix derived from the optimized structure.
    }\label{fig:CostFunction}
    
\end{figure*}

\begin{figure*}[t]
    \centering
    \includegraphics[width=0.70\linewidth]{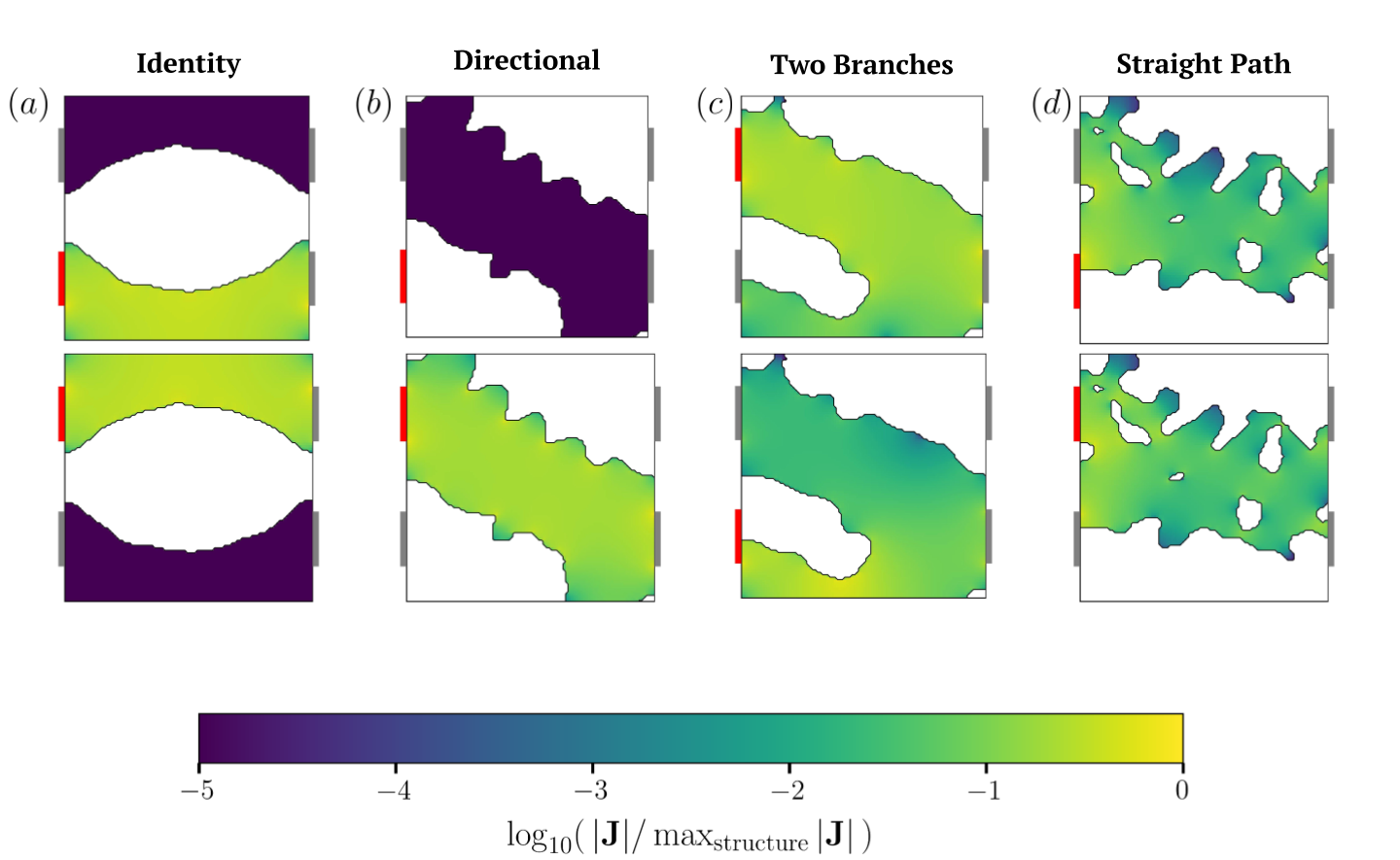}
\caption{Each panel shows the current normalized to maximum 1 in log scale. For each structure, we drive the two inports separately. Red bars mark driven (hot) ports; gray bars mark thermalized ports.
a)\emph{Identity}—heat splits symmetrically and exits through collinear ports, approximating an identity map.
b)\emph{Directional}—anisotropy in \(\mathbf{A}\) suppresses one branch and funnels heat along a single corridor (near-selector).
c)\emph{Two Branches}—two comparable channels divide the flux between outputs (operator with sizable off–diagonal weights).
d) \emph{Straight Path}—a reinforced backbone routes heat along a near-straight channel (strongly diagonal map). Across all cases, the spatially varying conductivity tensor \(\mathbf{A}\) sculpts low-resistance corridors that realize the desired boundary map \(\mathbf{M}\) while suppressing leakage.}\label{fig:heatFluxes}

\end{figure*}

We apply our inverse-design framework to synthesize a variety of target matrices $\mathbf{M}_{\mathrm{target}}$ with increasing structural and functional complexity. The optimized structures and their corresponding matrices are obtained after full binarization and removal of isolated islands (which only introduce noise since they have neglible flux). The units of the matrices $\mathbf{M}$ are mWK$^{-1}$, unless otherwise specified. The submatrices $\mathbf{A}$ have units W/K, where the effective thickness $\mathbf{z}$ are in $\mu$m. We will not specify units throughout the examples for readability. Furthermore, for visualization purposes, all the matrices are rounded to two decimal digits. The errors are reported in Tab.~\ref{table:Accuracy}. In this section, we attain matrices that can be represented by a single structure. We begin with the $2\times2$ identity matrix 
\begin{equation} \nonumber 
\mathbf{M}_{\mathrm{target}} = 
\begin{pmatrix}
1 & 0 \\[4pt]
0 & 1
\end{pmatrix},
\label{eq:Identity}
\end{equation}
which entails two separate simulations, one for each input port. Figure~\ref{fig:heatFluxes}(a) shows the convergence trajectory for increasing values of $\beta$. The optimized structure, shown in Fig.~\ref{fig:CostFunction}(a), is physically intuitive: connecting each input port directly to its corresponding output port. Fig.~\ref{fig:heatFluxes}(a) shows that the heat paths are two separate structures connecting one input and output at the same $y$-coordinate, consistent with the fact that $\mathbf{M}_{\text{target}}$ has no off-diagonal elements. The  computed matrix,
\begin{equation}\nonumber 
\mathbf{M} =
\begin{pmatrix}
0.99 & 0.00 \\[4pt]
0.00 & 1.01
\end{pmatrix},
\label{eq:ResultIdentity}
\end{equation}
is $<1\%$ from the target one. This matrix is decomposed as $\mathbf{M} = z_0\mathbf{A}(\rho_0)$, where $z_0$ = 27.6 $\mu$m and
\begin{equation}\nonumber
\mathbf{A}(\rho_0) =
\begin{bmatrix}
34.41 & 0.00 \\
0.00 & 34.77
\end{bmatrix}.
\end{equation}
The next example is the $2\times2$ directional matrix, which couples one input port with the opposite output port, producing a single off-diagonal element, 
\begin{equation}
\begin{aligned}\nonumber
\mathbf{M}_{\mathrm{target}} &=
\begin{pmatrix}
0 & 1 \\[4pt]
0 & 0
\end{pmatrix}
\end{aligned}.
\end{equation}
The computed matrix resembles the target one with error $~10^{-5}\%$. The optimized structure (Fig.~\ref{fig:CostFunction}(b)) exhibits a clean, heat channel between non-corresponding ports. The precision is again within $1\%$ precision. As shown in Fig.~\ref{fig:heatFluxes}(b), heat follows a single path from these two ports. For these two examples, it is extremely simple to devise the corresponding structures intuitively; thus, they serve as a validation for our framework. The next target matrix, which we refer to as the ``Two-Branch'' matrix, has two nonzero elements in a single row or column, corresponding to two parallel thermal channels. The target and output matrices are
\begin{equation}
\begin{aligned}
\mathbf{M}_{\mathrm{target}} &=
\begin{pmatrix}\nonumber
1.21 & 0.92 \\[4pt]
0 & 0
\end{pmatrix},
\qquad
\mathbf{M} &=
\begin{pmatrix}
1.20 & 0.91 \\[4pt]
0 & 0
\end{pmatrix}
\end{aligned},
\end{equation}
which are within $1\%$. The converged structure (Fig.~\ref{fig:CostFunction}(c) exhibits two well-defined parallel paths, and the final matrix achieves sub-percent accuracy. As shown in Fig.~\ref{fig:heatFluxes}(c) heat goes from each branch towards the same output port and each path has different distances and to the outport. The size of each path and their distances from the output port is what leads to the difference between the sizes of each entry. In our $100\times100$ design grid (filter radius $w=8$ pixels, port width $30$ pixels), the achievable ratio between branch conductances is typically limited to $\sim$3:1 due to smoothing and physical connectivity constraints. Next, we consider the ``Straight-Path'' matrix, where all elements are comparable in magnitude. The target and output matrices are
\begin{equation}
\begin{aligned}\nonumber
\mathbf{M}_{\mathrm{target}} &=
\begin{pmatrix}
1.21 & 1.50 \\[4pt]
1.18 & 1.31
\end{pmatrix},
\qquad
\mathbf{M} &=
\begin{pmatrix}
1.22 & 1.42 \\[4pt]
1.17 & 1.38
\end{pmatrix}
\end{aligned}.
\label{eq:StraightPath}
\end{equation}
The optimized design (Fig.~\ref{fig:CostFunction}(d) forms a uniform conductive slab, connecting all the input and output ports, corroborated by the flux patterns illustrated by Fig.~\ref{fig:heatFluxes}(d). In this case, the error is $~4\%$, which is higher than in the previous cases. This discrepancy is due to the fact that the distances between the input and output ports are not uniform, with pairs representing diagonal entries being closer. Therefore, our structures are naturally suitable for dominant diagonal matrices (assuming that all elements are nonzero). In this case, where all the entries have comparable magnitude, it becomes challenging to find optimal structures; matrices with dominant off-diagonal entries are out of reach for single-structure designs. In the next section, we show how we overcome this limitation by representing the target matrix with a combination of multiple structures, concurrently optimized.

\begin{figure*}[t]
    \centering
    \includegraphics[width=1\linewidth]{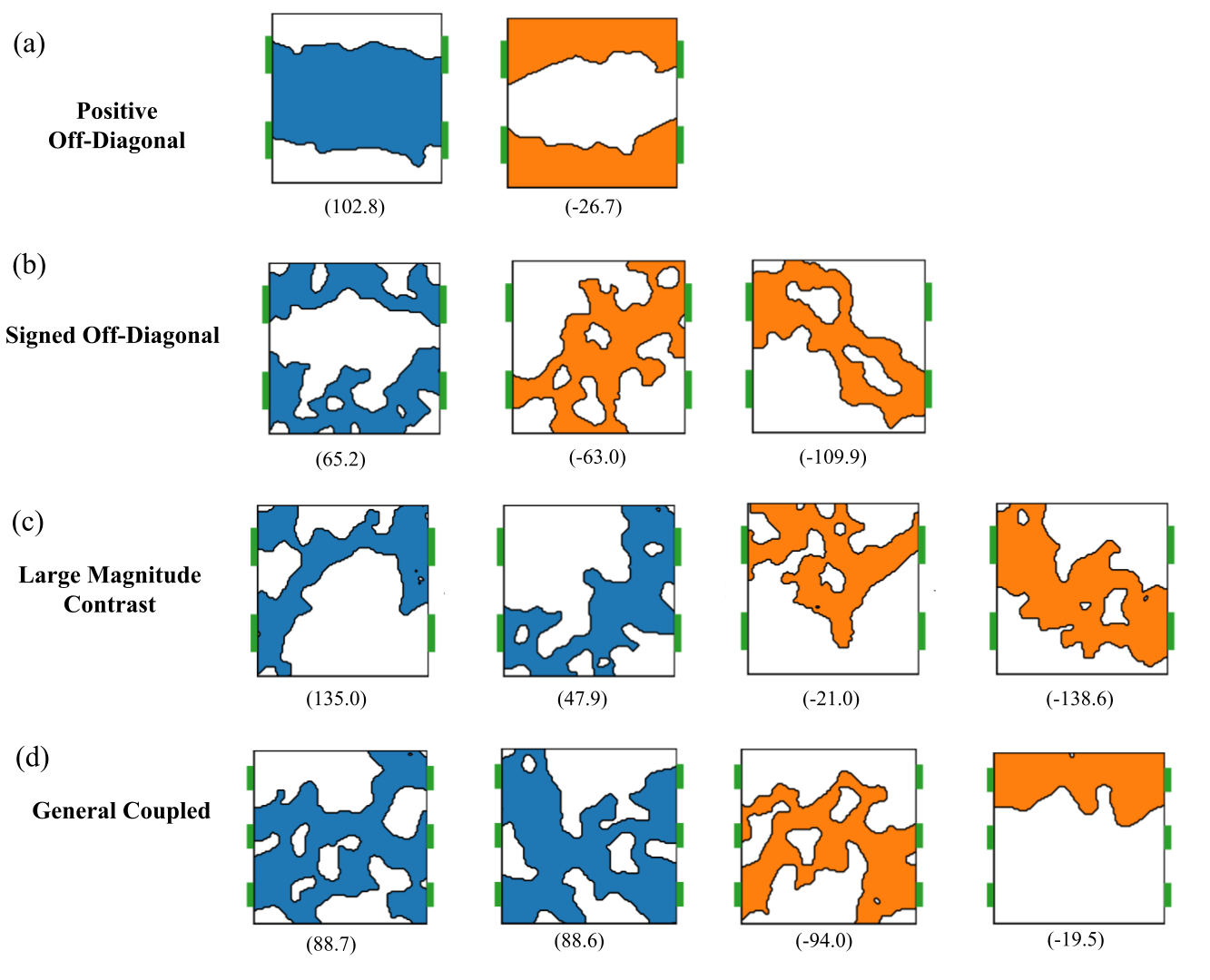}
    \caption{Each row shows the multi-structure representation of a given target matrix. Beneath each structure, the corresponding effective thickness $Z_i$ is shown. Blue and orange structures are associated with positive and negative $z_i$, respectively. (a) Positive-Off Diagonal ($2\times2$), (b) Signed Off-Diagonal ($2\times2$), (c) Large Magnitude Contrast ($2\times2$), and (d) General Coupled ($3\times3$).}
    \label{fig:multiple}
\end{figure*}

\section{Multiple-structure Designs}\label{sec:multiple}
In this section, we extend the range of representable matrices by increasing the number of structures, $K$. The first example is a $2\times2$ matrix with positive dominant off-diagonal entries (``Positive Off-Diagonal''). The target and output matrices are
\begin{equation}
\begin{aligned}
\mathbf{M}_{\mathrm{target}} &=
\begin{bmatrix}\nonumber
1.0 & 2.2 \\[4pt]
1.5 & 1.0
\end{bmatrix},
\qquad
\mathbf{M} &=
\begin{bmatrix}
0.99 & 2.20 \\[4pt]
1.51 & 1.01
\end{bmatrix},
\end{aligned}
\end{equation}
with an error below $1 \%$. The optimizer decomposes the output matrix in $K=2$ component
\begin{equation}\nonumber
\mathbf{M}_{\mathrm{target}}
\;\approx\;
\begin{aligned}[t]
&102.4\times 10^{-3}\times
\begin{pmatrix}
18.79 & 21.37 \\[4pt]
14.65 & 18.50
\end{pmatrix}
\\[6pt]
&-\,26.6 \times 10^{-3}\times
\begin{pmatrix}
35.19 & 0.00 \\[4pt]
0.00 & 33.37
\end{pmatrix}
\end{aligned}
\end{equation}
where the structures for each submatrix are shown in Fig.~\ref{fig:multiple}(a). Notably, the coefficient vector $\mathbf{z}=[102.4,-26.6]$ has one negative entry even though the target matrix has all positive entries. This result demonstrates the flexibility of our optimization pipeline and its ability to handle arbitrary matrices. The next example is a $2\times2$ matrix with dominant off-diagonal terms both positive and negative entries (``Signed Off-Diagonal''). The target matrix is
\begin{equation}
\begin{aligned}\nonumber
\mathbf{M}_{\mathrm{target}} &=
\begin{bmatrix}
1.0 & -3.0 \\[4pt]
-1.5 & 1.0
\end{bmatrix}
\end{aligned},
\end{equation}
which is almost identical to the output one, with an error below $10^{-5}\%$. In this case, an additional structure is needed ($K=3$), with $\mathbf{z}$ having  one positive entry and two negative entries. The optimized structures are shown in Fig.~\ref{fig:multiple}(b). The following matrix introduces elements differing by nearly an order of magnitude (''Large Magnitude Contrast''), requiring additional independent paths to maintain accuracy. The target and output matrices are
\begin{equation}
\begin{aligned}\nonumber
\mathbf{M}_{\mathrm{target}} &=
\begin{bmatrix}
0.5 & -5.0 \\[4pt]
2.0 & 1.0
\end{bmatrix},
\qquad
\mathbf{M} &=
\begin{bmatrix}
0.50 & -5.00 \\[4pt]
2.02 & 0.98
\end{bmatrix},
\end{aligned}
\end{equation}
which are within $1.2\%$. The optimzer requires an additional structure, thus $K=4$, with $\mathbf{z}$ having two positive and two negative entries. This example shows the limits of the dynamic range achievable in a fixed-resolution domain and the need for additional $K$ when matrix elements span multiple scales. The optimized structures are shown in Fig.~\ref{fig:multiple}(c). As the last example, we showcase our method for a fully-coupled $3\times3$ matrix with mixed amplitude (``General Coupled''). The target and output matrices are
\begingroup
\setlength{\arraycolsep}{4pt}
\renewcommand{\arraystretch}{1.05}
\[
\begin{aligned}
\mathbf{M}_{\mathrm{target}} &=
\begin{bmatrix}
0.42 & 0.22 & 0.97\\
0.58 & 0.54 & 0.36\\
0.88 & 0.64 & -0.54
\end{bmatrix},
\\[6pt]
\mathbf{M} &=
\begin{bmatrix}
0.41 & 0.20 & 0.96\\
0.57 & 0.53 & 0.36\\
0.89 & 0.62 & -0.54
\end{bmatrix}.
\end{aligned}
\]
\endgroup
which are within $<3\%$. For this example $K=4$ with $\mathbf{z}$ having two positive and two negative entries. The optimized structures are shown in Fig.~\ref{fig:multiple}(d). This case illustrates the generality of our method: complex multi-input/multi-output coupling—including negative and cross-interaction terms can be precisely realized with a small number of optimized metastructures. These results show the number of required structures scales with both the matrix dimension $D$ and sign composition:
\[
K \approx
\begin{cases}
D,  & \text{for fully positive or fully negative matrices},\\[6pt]
2D, & \parbox[t]{0.55\linewidth}{for signed matrices with \\[2pt] mixed positive and negative columns}.
\end{cases}
\]
This finding is supported by the fact that in the simplest case of a fully positive square matrix $\mathbf{M}_{\mathrm{target}} \in \mathbb{R}_+^{D\times D}$, each column can be implemented using a distinct $D$-branch structure. Hence, at most $K = D$ structures are sufficient to represent any such matrix, since each column defines an independent thermal mapping. The same argument holds for matrices with all negative entries. Hence, a generic $\mathbf{M}_{\mathrm{target}} \in \mathbb{R}^{D\times D}$ matrix needs at most $2D$ structures. However, we note that the grid resolution limits the contrast of the element within a single $D$-branch structures. For example, the grid used in this work ($100\times100$) won't be able to represent a D-branch with entries $[10^6,10^{-6}]$ in a single structure due to geometric limitations.  Hence, in such a case, we would need one structure for each element of the $D$-branch, giving the upper bound $K = 2D^2$, which holds for \emph{any} grid resolution.

\section{Application to notable matrices}\label{sec:applications}

\begin{figure*}[t]
    \centering
    \includegraphics[width=1.0\linewidth]{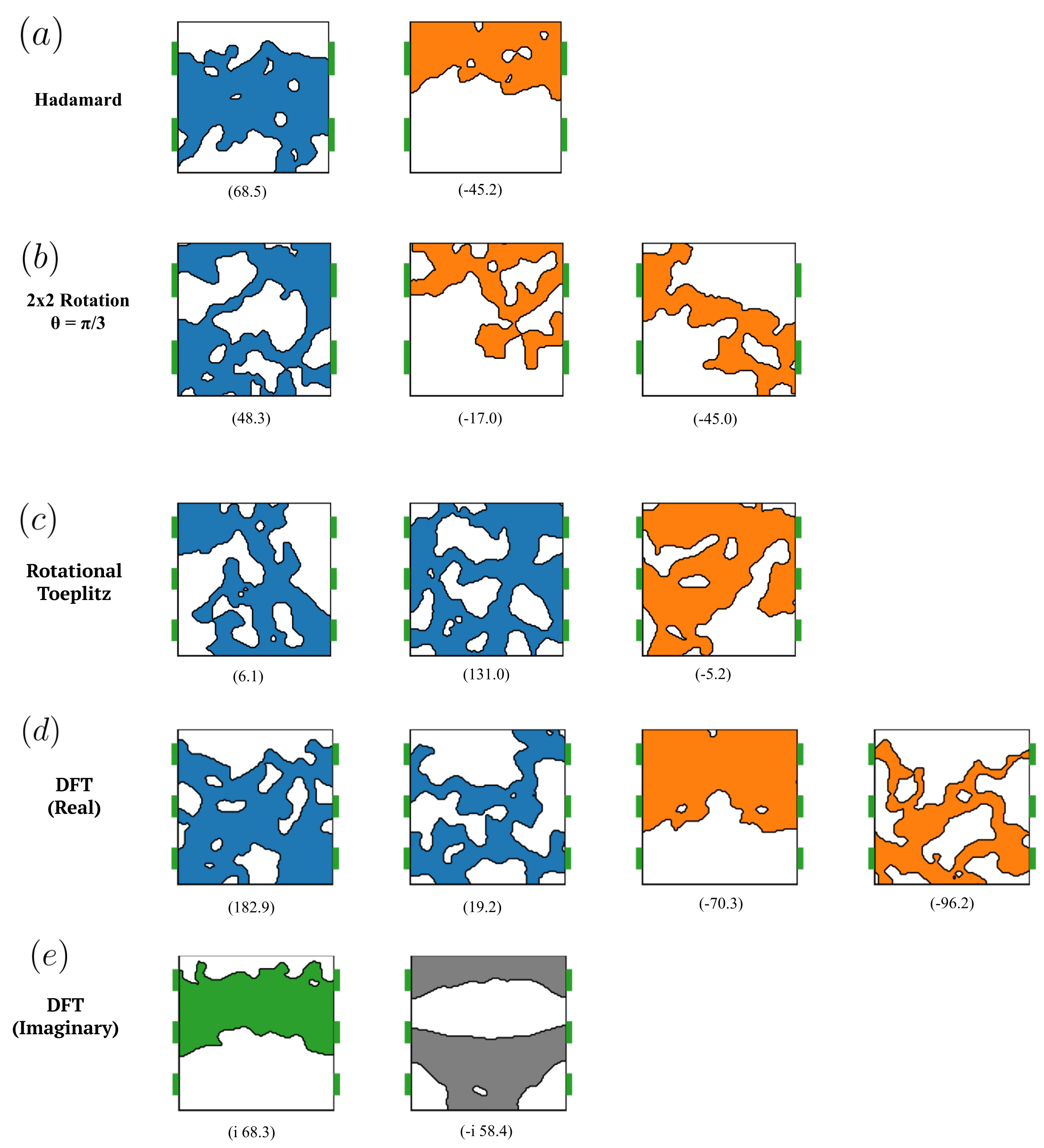}
\caption{
Each row shows the set of structures used for one target; the number beneath each structure gives its fitted thickness \(z_i\) (in \(\mu\mathrm{m}\)). In/out ports are shown in green on the left/right boundaries. Structures of different colors are post-processed separately by component: blue/orange denote \(+\)/\(-\) \emph{real} contributions, and green/gray denote \(+\)/\(-\) \emph{imaginary} contributions (DFT rows).\textit{(a)} Hadamard matrix (\(2\times2\)).
\textit{(b)} \(2\times2\) rotation with \(\theta=\pi/3\).
\textit{(c)} Rotational Toeplitz (\(3\times3\)).
\textit{(d)} \(3\times3\) Discrete Fourier Transform — real part.
\textit{(e)} \(3\times3\) Discrete Fourier Transform — imaginary part.}
    \label{fig:AllKdesigns}
\end{figure*}

In many practical applications—such as analog signal processing, filtering, and neuromorphic computation—the matrices of interest are known in advance, allowing these metastructures to serve as reusable physical computing elements. In 
the following, we highlight examples of important matrices realized by our method, together with their physical realizations and the corresponding $K$ values. We begin with the Hadamard matrix, which is a canonical orthogonal transform used in image processing and quantum logic. The target and computed matrices are
\begin{equation}
\begin{aligned}\nonumber
\mathbf{M}_{\mathrm{target}} &=
\begin{bmatrix}
1 & 1 \\[4pt]
1 & -1
\end{bmatrix}
\,,\qquad
\mathbf{M} &=
\begin{bmatrix}
1.00 & 1.00 \\[4pt]
1.00 & -1.01
\end{bmatrix}
\,,
\end{aligned}
\end{equation}
feature a deviation within $1\%$. The needed number of structures (Fig.~\ref{fig:AllKdesigns}(a)) is $K=2$. Next, we consider a generic 2×2 rotational matrix. For a rotation angle $\theta=\pi/3$ the target and computed matrices are
\begin{equation}
\begin{aligned} \nonumber
\mathbf{M}_{\mathrm{target}} &=
\begin{bmatrix}
0.5 & -0.866 \\[4pt]
0.866 & 0.5
\end{bmatrix},\,\,\,\,
\mathbf{M} =
\begin{bmatrix}
0.473 & -0.864 \\[4pt]
0.867 & 0.515
\end{bmatrix},
\end{aligned}
\end{equation}
where we use $K=3$. In this case, the error is $3\%$. The optimized structures are shown in Fig.~\ref{fig:AllKdesigns}(b). The next example is a ($3\times3$) rotational Toeplitz matrix, which encodes translationally invariant operations such as convolution. We demonstrate that such a system form can be implemented using $K=3$ structures (Fig.~\ref{fig:AllKdesigns}(c)). The target and computed matrices
\begin{equation}
\begin{aligned}\nonumber
\mathbf{M}_{\mathrm{target}} &=
\begin{bmatrix}
1.0 & 0.7 & 0.4 \\[4pt]
0.7 & 1.0 & 0.7 \\[4pt]
0.4 & 0.7 & 1.0
\end{bmatrix},\,\,\,
\mathbf{M} &=
\begin{bmatrix}
1.01 & 0.68 & 0.39 \\[4pt]
0.72 & 0.99 & 0.70 \\[4pt]
0.41 & 0.70 & 1.01
\end{bmatrix},
\end{aligned}
\end{equation}
which are within $~3\%$. The last application is the Discrete Fourier Transform (DFT), which underpins frequency-domain analysis in optics and signal processing. For $D=3$, the corresponding matrix reads
\begin{equation}
\mathbf{M}_{\text{target}} \;=\;
\begin{bmatrix}\nonumber
1 & 1 & 1 \\
1 & -0.5 & -0.5 \\
1 & -0.5 & -0.5
\end{bmatrix}
\;+\;
i\begin{bmatrix}
0 & 0 & 0 \\
0 & -0.87 & 0.87 \\
0 & 0.87 & -0.87
\end{bmatrix},
\end{equation}
which has an imaginary component. We decompose the target matrix into four contributions
\begin{equation}\nonumber
\mathbf{M} = \mathbf{M}^{(+)} - \mathbf{M}^{(-)} + i\big(\mathbf{M}^{(+i)} - \mathbf{M}^{(-i)}\big),
\end{equation}
where $\mathbf{M}^{(+)}$ and $\mathbf{M}^{(-)}$ denote the positive and negative real conductance maps, respectively. Similarly, $\mathbf{M}^{(+i)}$ and $\mathbf{M}^{(-i)}$ stand for the positive and negative imaginary parts. In this case, we have $K=4$ structures to encode the real part and $K=2$ to represent the imaginary one, yielding the output matrix
\begin{equation}
\mathbf{M} = 
\begin{bmatrix}\nonumber
1.01 & 0.95 & 1.05 \\
1.00 & -0.50 & -0.50 \\
1.01 & -0.53 & -0.52
\end{bmatrix}
\;+\; i
\begin{bmatrix}
0.00 & 0.00 & 0.00 \\
0.00 & -0.85 & 0.85 \\
0.00 & 0.87 & -0.89
\end{bmatrix},
\end{equation}
with errors of $3.55\%$ and $1.13\%$ for the real and imaginary part, respectively. The optimized structures  for the real component are shown in Fig.~\ref{fig:AllKdesigns}(d) and the imaginary is shown in Fig.~\ref{fig:AllKdesigns}(e).

\section{Performances}\label{sec:perfomances}

The bandwidth of our structures is limited by thermal diffusion. As detailed in Sec.~\ref{sec:bandwidth}, for a lateral dimension of $L = 100~\mu\text{m}$, the \emph{settling} time $\tau_s \approx 83.7\mu$s, corresponding to a bandwidth of $\Delta f \approx 1.9~\text{kHz}$. The energy needed for computation can be estimated as
\begin{equation}
E_{\text{op}} \;\approx\; \tau_s\,\|\mathbf{M}\|\,\Delta T,
\end{equation}
where $\|\mathbf{M}\|$ is the effective conductance (W/K), and $\Delta T$ is the applied temperature difference (K). For example, if $\|\mathbf{M}\|\!\sim\!10^{-3}~\text{W/K}$ and $\Delta T = 1$\,K, then $E_{\text{op}} \approx 83.7~\text{nJ}$. When benchmarking MVMs, one typically reports the energy per multiply–accumulate (MAC) operation. Assuming a $3\times 3$ matrix with nine MAC operations, the energy consumption is $9.3\text{nJ/MAC}$. To put these numbers in context, electronic crossbars perform MVMs in the MHz–GHz range with fJ–pJ/MAC, ~\cite{sebastian2020memory,ielmini2018inmemory}, while photonic accelerators operate in the GHz–THz regime with energy consumpion in the pJ/MAC range~\cite{shen2017coherent,wetzstein2020optical}. Although our structures are less performant than photonic or memristive systems, their fundamental advantage lies in the fact that the physical signal \emph{is} the computation itself, eliminating the need for actively powered matrix operations. This makes our platform energy-passive, non-volatile, and naturally integrable with microelectronic environments where heat is already a byproduct of computation.

Power efficiency is a key metric for analog computing. In our case, potential losses are due to thermal radiation and thermal-crosstalk, the latter involving air-assisted heat conduction. However, as detailed in Sec.~\ref{sec:energy}, both contributions are neglibile compared to the output power for reasonable applied tempeature differences. In Sec.~\ref{sec:snr}, we also estimate the signal-to-noise ratio (SNR) arising from thermal fluctuations, finding  SNR $\approx10^6$ at room temperature. Therefore, neither energy loss nor SNR constitutes a primary limitation of our methodology. Lastly, we assess the validity of assumption of the linear regime, which imposes an upper bound on the range of the applied difference of temperature. In Sec.~\ref{sec:nonlinear}, we perform
nonlinear steady-state, heat conduction simulations for one of the structures considered in this work. Using silicon with a base temperature of $T_0 = 300$~K, we find that the accuracy of the MVM operation remains within $1\%$ and $5\%$ for applied temperature differences up to 25K and 100K, respectively. These results demonstrate that our structures can reliably perform MVMs across a temperature range relevant to key applications, such as hot spots in microelectronics. Lastly, we note that our inverse-design framework is robust against initialization. To demonstrate this aspect, we run 10 simulations for one of the sructures discussed above (precisely, the ``Two-Branch'' one), using different random initial guesses. We found that the performance of this test depends on the $\beta$-schedule, where doubling the maximum number of iterations for each epoch yields errors below $1\%$ in 100 $\%$ of the cases.

\section{Validity of the Diffusive Regime}\label{sec:ballistic}

Lastly, we note that simulations are based on the standard heat conduction equation, and thus nondiffusive transport such as phonon-size scattering~\cite{chen2021non} are not captured. Consequently, we need to ensure that the characteristic size of the structure is much larger than the mean-free-path of the dominant phonon, which in Si at room temperature is $\Lambda \approx100 $nm~\cite{esfarjani2011heat}. In the three-field approach of topology optimization, the feature size is of the order of the conic filter radius, which in our case is $\tilde R = 8 \mu \textrm{m} >> \Lambda$; therefore, we are in a regime where we can safely ignore phonon size effects. We note that a more rigorous approach to setting the size of the feature is by using~\emph{geometric constraints}~\cite{zhou2015minimum}. In future work, we plan to add these constraints for both ensuring that the diffusive regime is valid when we use Fourier's law as the physics engine and for ensuring manufacturability. Lastly, the limitation on the minimum lengthscale indirectly imposes the size of the overall system, $L$. In fact, to encode a wide range of matrices, our design must be much larger than the feature size. In our experiments, we found that $L\approx 100/8 \tilde R = 100 \mu$m. Extending our methodology to the nanoscales entails solving the phonon Boltzmann transport equation (BTE)~\cite{Ziman2001}. Recently, we have developed a differentiable phonon BTE solver, which was integrated into OpenBTE~\cite{romano2021openbte}. We applied this tool to the design of metastructures with effective thermal conductivity~\cite{romano2022inverse}, which is a problem with the same complexity as the one tackled in this work. As a possible future direction, we plan to employ OpenBTE to design nanostructures for MVMs, thus extending our proposed methodology to the nanoscales.

\section{Discussions}\label{discussions}

In the previous section, we showcase our methodology for a variety of structures achieving high accuracy in all cases. Despite these promising results, several constraints were identified. First, as the size of the matrix increases, the complexity of the conductive topology increases, requiring a finer spatial discretization and higher computational cost. Second, the physical arrangement of ports introduces intrinsic asymmetries: ports located centrally have more balanced coupling paths than those near the edges. Additionally, the finite spacing between contacts ($L$) imposes a limit on the response speed. Future work will push towards: (i) automatic discovery of the number of needed matrices, (ii) adding lengthscale constraint to ensure simpler shapes conducing the heat,  (iii) scalability to higher dimensional matrices through tiling of sub-
matrices, and (iv) harnessing the temperature dependence of thermal conductivity to realized nonlinear transformations.

As detailed in Sec.~\ref{sec:energy}, fluctuations in the surrounding air do not significantly affect the measured heat flux in our structures. The thermal conductivity of crystalline silicon is approximately $6\times10^3$ times larger than that of air, and at the sub-micrometer scales considered here, heat transfer is conduction-dominated while convective effects are strongly suppressed. Consequently, embedding the optimized metastructures in air does not degrade their performance. These geometries are already compatible with standard CMOS fabrication processes—e.g., shallow trench isolation, air-gap interconnects, and phononic/photonic crystals. These can be  manufactured at wafer scale without the need for specialized vacuum environments~\cite{CMOSAirGap,PhononicCrystalReview}.

\section{Conclusions}\label{sec:conclusions}
We introduced a novel platform for matrix–vector multiplication (MVM), wherein heat flow itself performs the linear operation. By combining a differentiable, JAX-based steady-state solver with density-based topology optimization, we inverse-designed planar metastructures whose geometry encodes a target conductance map. We demonstrated accurate implementations of representative $2{\times}2$ and $3{\times}3$ operators---including Hadamard, Toeplitz, rotational, and discrete Fourier transform matrices. Across all cases, the recovered matrices closely matched their targets, validating that heat transport can be engineered to implement useful linear transforms with high fidelity. While such structures will not replace electronic or photonic accelerators for high-throughput workloads, they open a complementary regime in which computation and dissipation coincide. Future directions include extending the platform to nanoscale heat transport, where diffusive behavior breaks down, and to larger matrix dimensions. These results establish a proof of concept for thermal analog computing architectures in which heat itself serves as the computational signal.

\section{Data availibility statement}
The data that support the findings of this article are openly available~\cite{silva_2025_18048664}.

\section{Acknowledgements}

We thank the MIT Undergraduate Research Opportunities Program.

\appendix


\section{Barrier Function}\label{sec:optimization}

\begin{figure}[t]
    \centering
    \includegraphics[width=\linewidth]{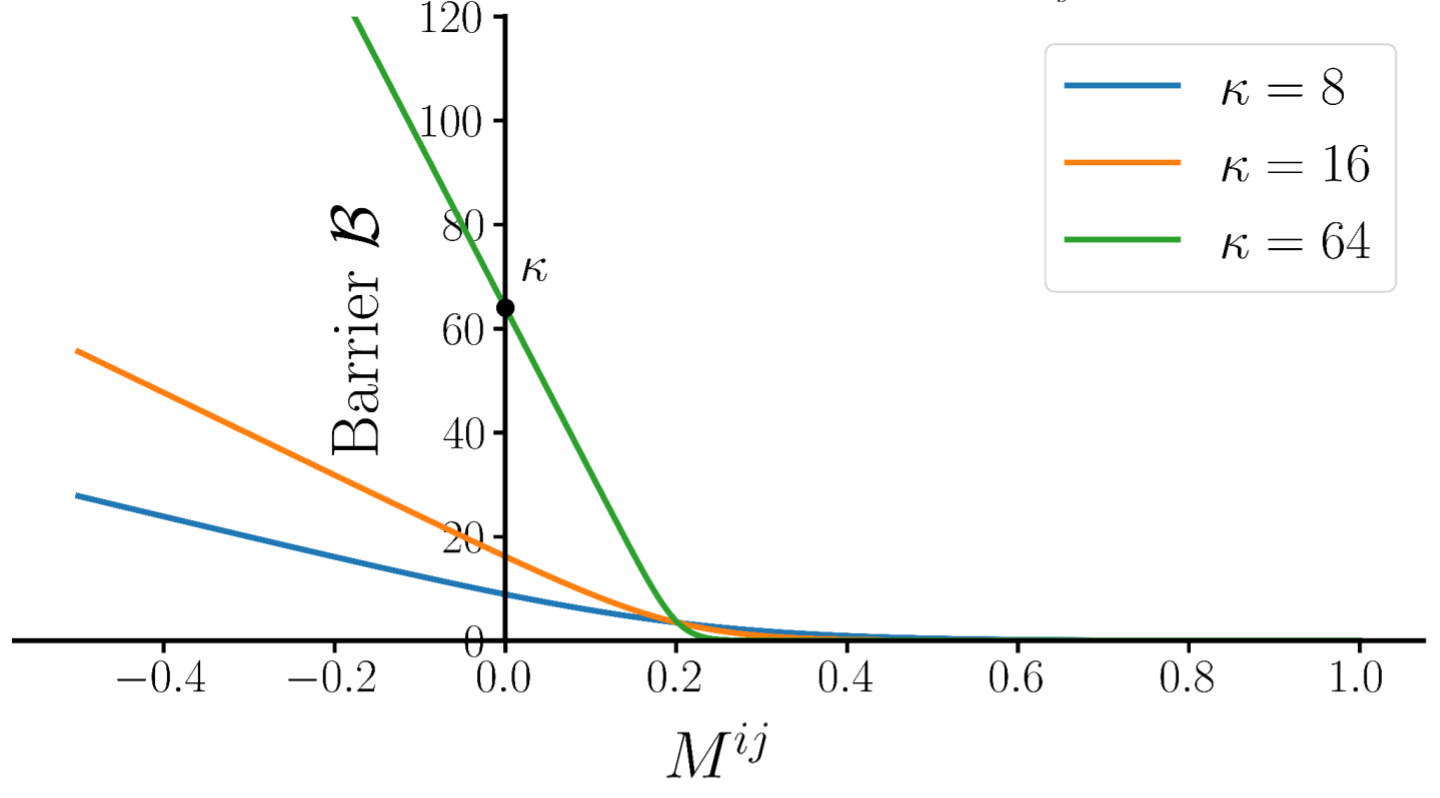}
    \caption{
        Softplus-based barrier function for different curvature parameters 
        $\kappa_b \in \{8,16,64\}$, evaluated for a normalized target magnitude 
        $|M_{ij}^{\mathrm{target}}| = 1$, with threshold $\tau = 0.2$ and regularizer 
        $\varepsilon = 10^{-2}$.  
        For $X$ exceeding the threshold ($X > \tau$), the penalty approaches zero, while 
        for undershoot values ($X<\tau$) the barrier becomes approximately linear with slope 
        proportional to $\kappa_b$.  
        Larger $\kappa_b$ yields a sharper transition and stronger corrective gradients, 
        matching the behavior used in our topology-optimization pipeline.
    }
    \label{fig:SoftPlus}
\end{figure}

To prevent individual matrix elements from undershooting their target values during the inverse-design optimization, we introduce a differentiable softplus-based barrier. The barrier for a single matrix entry is defined as
\begin{equation}\label{eq:Barrier}
    \mathcal{B}
    =  
    \sum_{i,j}
    \kappa_b \cdot 
    \mathrm{softplus}\!\left[
        \kappa_b \left(
        \tau -
        \frac{\mathrm{sign}(M^{ij}_{\mathrm{target}})\,M^{ij}}{|M^{ij}_{\mathrm{target}}|+\varepsilon}
        \right)
    \right],
\end{equation}
and depends on four quantities:  
(i) the admissible fractional undershoot $\tau$,  
(ii) a small regularizer $\varepsilon$ to avoid division by zero,  
(iii) the local ratio between the achieved and desired matrix values, and  
(iv) a curvature parameter $\kappa_b$ controlling the steepness of the penalty. When the normalized ratio 
\begin{equation}\nonumber
\frac{\mathrm{sign}(M^{ij}_{\mathrm{target}})\,M^{ij}}{|M^{ij}_{\mathrm{target}}|+\varepsilon}
\end{equation}
exceeds the threshold $\tau$, the barrier contribution becomes negligible;  
when the ratio falls below this threshold, the penalty grows approximately linearly, with slope proportional to $\kappa_b$. The softplus barrier used in our optimization is defined in Eq.~\ref{eq:Barrier}. The barrier provides a smooth but strong gradient signal that pushes each $M^{ij}$ back toward its desired value. In this work, we choose $\tau = 0.2$, $\varepsilon = 10^{-2}\,\|M_{\mathrm{target}}\|_F$,  $\kappa_b = \min(\beta,\,256)$. At the beginning of the optimization, $\beta$ (and therefore $\kappa_b$) is small, so the barrier remains weak and does not restrict geometry exploration.  
As the projection sharpens, the barrier becomes stronger, ensuring that any matrix element trending below its target is corrected even in the late stages of topology refinement. Saturating $\kappa_b$ at $256$ prevents numerical overflow while maintaining a steep barrier enough to reliably correct the undershoot. Figure~\ref{fig:SoftPlus} shows the one-dimensional barrier response as a function of $M^{ij}$ for various $\kappa$. The curve corresponds to the values of the barrier of a single matrix entry appearing in Eq.~\ref{eq:Barrier} for a unit-magnitude target matrix entry.

\section{Bandwidth}\label{sec:bandwidth}

\begin{figure}[t]
    \includegraphics[width=0.9\linewidth]{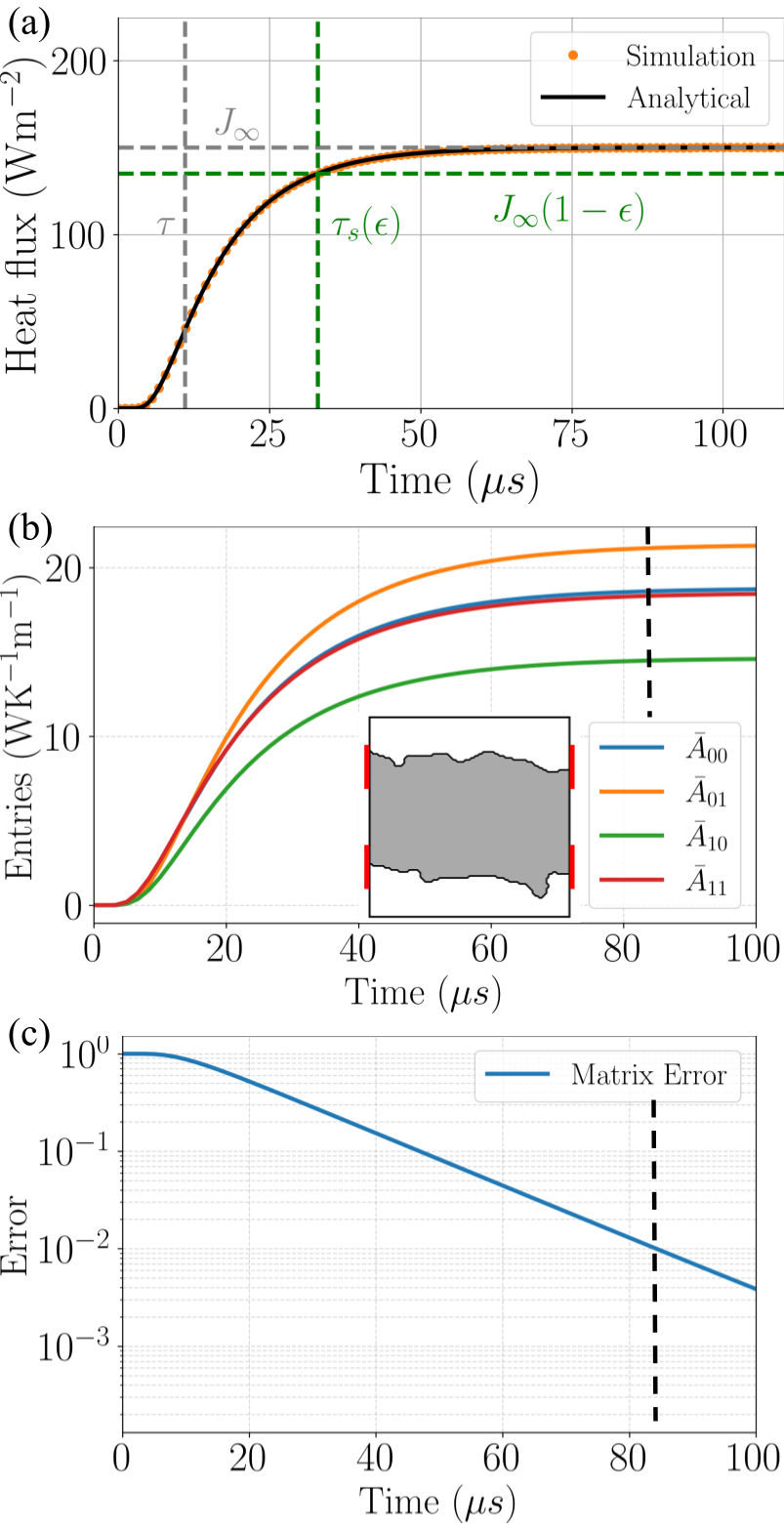}
\caption{
(a) Transient response for a simple homogeneus case, superimposed to the analytical solution from Eq.~\ref{eq:flux}. Vertical lines marking $t = \tau$ and $t = \tau_s(0.1)$ are also shown. (b) Transient response from the optimized structure implementing the positive contribution of the ``Positive Off-Diagonal'' matrix. The plotted values are the time evolution of the recovered matrix entries $\mathbf{\bar{A}}(t)$ from the output heat flux after a unit‐step excitation. (c) Accuracy (log scale) defined in Eq.~\ref{eq:et}. Both (b) and (c) also include a vertical line for $t = \tau_s(10^{-3})\approx 83.7 \mu$s.}
\label{fig:transient}
\end{figure}

The bandwidth of our structure is limited by the thermal relaxation time, which can be estimated by
\begin{equation}\label{eq:bandwidth}
\Delta f\approx\frac{1}{2\pi\tau_s},
\end{equation}
where $\tau_s$ is the \emph{settling} time. To estimate $\tau_s$, we consider a uniform system with the same $xy$-plane dimensions and material as the structures used throughout this work; that is, the in-plane domain is a square with side $L=100~\mu$m, and the thermal conductivity is that of silicon at 300 K, i.e., $\kappa = 150$ W$^{-1}$ K$^{-1}$. We assume an infinite thickness. For transient simulations, we also have to define the heat capacity, $C$, which for Si at 300 K is $1.63\times 10^6$ Jm$^{-3}$K$^{-1}$, giving a thermal diffusivity of $\alpha = \kappa/C =  0.92$  cm$^2$s$^{-1}$. We apply a temperature difference $\Delta T = 1$ K across the $x$-axis. This scenario has an analytical solution,
\begin{equation}\label{eq:flux}
J(t) = J_{\infty}\!\left[
1 + 2 \sum_{n=1}^{\infty} (-1)^n
\exp\!\left( -n^2\frac{t}{\tau} \right)
\right],
\end{equation}
where $J_{\infty} = -\,\kappa\,\Delta T / L$, and $\tau = (L/\pi)^2/\alpha\approx 11 \mu$s is the relaxation time for the first mode. We define the settlement time $\tau_s(\epsilon)$ as the time required by the system to have the flux $J(\tau_s) \approx J_\infty\left(1-\epsilon\right)$. From Eq.~\ref{eq:flux}, we have $\tau_s(\epsilon) \approx \tau \text{ln}(2/\epsilon)$. In this work, we chose $\epsilon = 0.1\%$, giving $\tau_s \approx \gamma\tau = 83.7 \mu$s. The corresponding bandwidth is evaluated by Eq.~\ref{eq:bandwidth}, which yields 
\begin{equation}\label{eq:bandwidth2}
\Delta f = \frac{\alpha\pi}{2 L^2 \gamma}\approx 1.9\,\text{KHz}.
\end{equation}
Next, we study the transient effect of a realistic structure, by directly solving the time-dependent heat conduction equation
\begin{equation}\label{eq:transient}
C\frac{\partial T}{\partial t} = \nabla \cdot\kappa \nabla T.
\end{equation}
To this end, we first test the transient solver in the uniform case discussed above, achieving excellent agreement with Eq.~\ref{eq:flux} (Fig.~\ref{fig:transient}(a)). In solving Eq.~\ref{eq:transient}, we choose the Runge-Kutta algorithm ``Tsit5'' time-stepping method, implemented in Diffrax~\cite{kidger2021on}. We now move on to simulating the transient response of one of the structures identified in this work, precisely the one encoding the positive contribution of the ``Positive Off-Diagonal'' matrix. We refer to the corresponding submatrix as $\mathbf{A}$. We solve Eq.~\ref{eq:transient} together the boundary conditions from Eq.~\ref{eq:fourier} with an applied temperature of $\Delta T = $ 1 K for each excitation. As shown in Fig.~\ref{fig:transient}(b), the time-dependent reconstructed matrix $\bar{\mathbf{A}}(t)$ is very close to its steady-state value. Figure~\ref{fig:transient}(c) plots the time-dependent deviation from the steady-state matrix, calculated as
\begin{equation}\label{eq:et}
\epsilon(t) = ||(\bar{\mathbf{A}}(t)-\mathbf{A} ) \odot 1/\mathbf{A}||_F.
\end{equation}
We see that for $t = \tau_s$, the error is $\epsilon(t) <  1\%$. These results corroborate considering $\tau_s(10^{-3})$ as a suitable settling time.

\section{Energy efficiency}\label{sec:energy}

We provide an estimate of the power loss via conduction through air and radiation, which can also potentially trigger thermal crosstalk among the structures. Similarly to Sec.~\ref{sec:snr}, we consider a uniform slab of Si at $T_0 = $ 300 K and in-plane geometry of a square with side  $L = 100\mu$m. In this example, we assume a finite thickness $L_z = 0.5 L$, which results in a reasonable shape factor.  One side of the structure is thermostatted at $T_0$ = 300 K, while the other one is kept fixed at $T =$ 400 K, thus resulting in a difference of applied temperature of $\Delta T = 100$K. The output power is then
\begin{equation}\label{power_output}
P_{\textrm{out}} = \kappa \Delta T L_z = 750 \text{mW},
\end{equation}
with $\kappa$ being the thermal conductivity of Si given in Sec.~\ref{sec:bandwidth}. The power lost to radiation through the outer surface is
\begin{equation}\nonumber
 P_{\text{rad}} = \sigma4L^2\left(T^4-T_0^4\right)\approx 0.05\text{mW},
\end{equation}
where $\sigma$ is the Stefan-Boltzmann constant, giving $P_{\text{rad}}/P_{\text{out}}\approx 5.3\times 10^{-5}$, which is neglibile especially considering that in practical applications the applied difference of temperatue is generally less than $100$ K. Heat can also be lost as a result of thermal conduction through air. To estimate this effect, we consider two structures stacked along $z$ (Fig.~\ref{fig:MVM}) with separation $h = L$. We assume that one structure is at uniform temperature $T$ and the other one at $T_0$. The power transferred from one structure to the other is then 
\begin{equation}\nonumber
P_{\text{cross}} =\kappa_{A}L^2\Delta T/h = 0.52 \text{mW}
\end{equation}
with $\kappa_A = 0.026$ W m$^{-1}$K$^{-1}$ being the thermal conductivity of air. The relative loss is $P_{\text{cross}}/P_{\text{out}}\approx 7\times 10^{-5}$, thus negligible compared to the output power.

\section{Signal-to-noise ratio}\label{sec:snr}

Heat flux exhibits intrinsic thermal fluctuations, which  is a potential source of noise. To compute the signal-to-noise (SNR) ratio, we consider the same uniform system analyzed in Sec.~\ref{sec:bandwidth}, but with finite thickness $L_z = 0.5 L$, which results in a reasaonble shape factor. In presence of noise, the measured output power can be expressed as 
\begin{equation}\label{eq:timepower}
P_{\text{out}}(t)=\bar P_{\text{out}}+\delta P_{\text{out}}(t).
\end{equation}
where the average power is $\bar P_{\text{out}} = G\Delta T$, with $G = \kappa L_z$. ($\kappa$ is the thermal conductivity of Si defined in Sec.~\ref{sec:bandwidth}).
Equation~\ref{eq:timepower} has zero-mean fluctuations $\langle \delta P_{\text{out}}(t)\rangle=0$, and variance \begin{equation}\label{eq:variance}
\mathrm{Var}(P_{\text{out}})=\mathrm{NEP}^2\,\Delta f,
\end{equation}
where the $\Delta f$ is defined in Eq.~\ref{eq:bandwidth}, and NEP is the thermal fluctuation noise-equivalent power~\cite{mather1982bolometer},
\begin{equation}\label{eq:NEP}
\mathrm{NEP}=\sqrt{4k_B \bar T^{2} G}\qquad [\mathrm{W}/\sqrt{\mathrm{Hz}}],
\end{equation}
In Equation~\ref{eq:NEP}, $k_B$ is the Boltzmann constant, and $\bar T=(T+T_0)/2$ is the mean temperature. Combining Eqs.~\ref{eq:variance}–\ref{eq:NEP}, we have the SNR
\begin{align}
\mathrm{SNR}\label{eq:snr}
&=\frac{\bar P_{\text{out}}}{\sqrt{\mathrm{Var}(P_{\text{out}})}}
=\Delta T\,\sqrt{\frac{G}{4k_B \bar T^{2}\Delta f}}
=\nonumber \\ &= \frac{\Delta T}{\bar T}\frac{L}{L_z}\,\sqrt{\frac{\gamma C_{\text{tot}}}{2\pi k_B}} \approx 1.78\times 10^6
\end{align}
where $C_{\text{tot}}=C L^{2}L_z$ is the total heat capacity of the slab. This SNR is computed by assuming a small $\Delta T = 1$ K, which is generally lower than that in practical applications.

\section{Sensitivity to temperature-dependent thermal conductivity}\label{sec:nonlinear}

\begin{figure}[t]
    \includegraphics[width=0.8\linewidth]{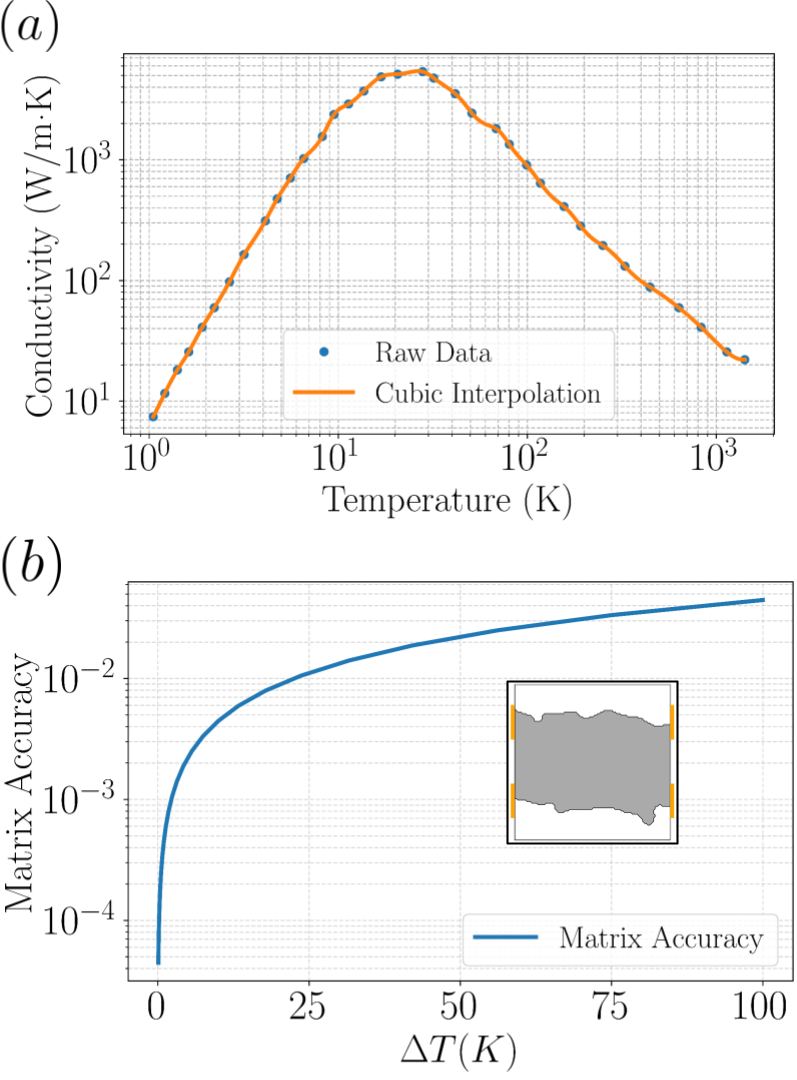}
\caption{
(a) Temperature-dependent thermal conductivity of silicon. Blue points show measured data from Ref.~\cite{Ziabari2010AdaptivePowerBlurring}, and the orange curve is a cubic spline interpolation used in our nonlinear simulations.  
(b) Linearization error $\epsilon(\Delta T)$ (Eq.~\ref{response}) of the inverse-designed structure as a function of input temperature excursion $\Delta T$ from the base temperature $T_0=300\,$K. 
The inset shows the geometry used in the test.  
}\label{fig:nonlinear}
\end{figure}

The proposed topology optimization framework assumed temperature-independent thermal conductivity. In this section, we assess the range of validity of this assumption. To this end, we first solve the nonlinear heat conduction equation
\begin{equation}\nonumber
\nabla \cdot \left[\kappa(T) \nabla T \right]= 0
\end{equation}
over a uniform domain with side L = $100 \mu$m. The underlying material is Si, with temperature-dependent thermal conductivity $\kappa(T)$ obtained by a polynomial fit of the experimental data from Ref.~\cite{Asheghi1998_Si_kappa_film} (Fig.~\ref{fig:nonlinear}(a)). We apply a temperature $T_0 = $ 300 K on the left and $T = $ 350 K on the right. The computed power density is in excellent agreement ($<1\%$) with the analytical solution 
\begin{eqnarray}\label{flux}
P &=& - \int_{T_0}^{T}\kappa(T)dT \approx 6.8 \text{mW}{\mu} \text{m}^{-1}.
\end{eqnarray}
As an example, we now calculate the nonlinear response of the structure representing the positive contribution of the matrix ``Positive Off-Diagonal.'' The output ports are kept at $T_0 = $ 300 K, and the vector of the input temperature can take large values, takking the form 
\begin{equation}\nonumber
\mathbf{T}_{\text{in}} = T_0 + \Delta T\boldsymbol\xi,
\end{equation}
where $\boldsymbol \xi $ is a one-hot vector and $\Delta T$ is the applied difference of temperature. Given the parameters $\boldsymbol \theta$, the calculated output is
\begin{equation}\label{response}\nonumber
\mathbf{P}_{\text{out}} = f_{\boldsymbol\theta}(\mathbf{T}_{\text{in}}),
\end{equation}
with $f_{\boldsymbol \theta}$ being the function response of the optimized structure. Using Eq.~\ref{response}, we reconstruct the \emph{effective} matrix $\mathbf{\bar A}^{\text{eff}}$ using the same procedure as in the main text. Finally, we define the linearization error as a function of $\Delta T$
\begin{equation}\nonumber
\epsilon(\Delta T) = ||(\mathbf{\bar A}^{\text{eff}}(\Delta T)-\mathbf{A}) \odot 1/\mathbf{A}||_F.
\end{equation}
which measures the mismatch between the effective operator and the ideally linear target matrix $\mathbf{A}$. As shown in Fig.~\ref{fig:nonlinear}(b), the MVM remains highly accurate (error $\sim1\%$) for applied temperature differences up to 25~K, while the deviation remains below 5$\%$ for differences up to $100$~K. The admissible temperature range therefore depends on the error tolerance required by the intended application, but for moderate input excursions, the linear approximation adopted in this work remains well justified.

\bibliography{biblio}  

\end{document}